\documentclass[prc,showpacs,twocolumn,amsmath,amssymb]{revtex4}
\usepackage{graphicx}
\usepackage{dcolumn}
\usepackage{bm}
\usepackage{booktabs}
\usepackage{color}
\begin{document}

\title{Effects of unconventional breakup modes on incomplete fusion of weakly bound nuclei}

\author{Alexis Diaz-Torres and Daanish Quraishi}

\affiliation{Department of Physics, University of Surrey, 
Guildford GU2 7XH, UK}

\date{\today}

\begin{abstract}
The incomplete fusion dynamics of $^6$Li + $^{209}$Bi collisions at energies above the Coulomb barrier is investigated. The classical dynamical model implemented in the {\sc platypus} code is used to understand and quantify the impact of both $^6$Li resonance states and transfer-triggered breakup modes (involving short-lived projectile-like nuclei such as $^8$Be and $^5$Li) on the formation of incomplete fusion products. Model calculations explain the experimental incomplete-fusion excitation function fairly well, indicating that (i) delayed direct breakup of $^6$Li reduces the incomplete fusion cross-sections, and (ii) the neutron-stripping channel practically determines those cross-sections.
\end{abstract}


\maketitle

\emph{Introduction.} The physics of reactions induced by exotic beams is at the core of many experiments at new generation facilities. Understanding fusion dynamics of light weakly bound nuclei is very important for characterizing astrophysical reactions critical for element creation \cite{Wiescher1}. The breakup mechanism of weakly bound nuclei is vital to understanding the dynamics of fusion as well as the consequences of breakup \cite{Back1,Canto1}. Broadly there are three possibilities of reaction processes upon breakup of the light nucleus. The first is where no fragments are captured and is termed \emph{no-capture breakup} ({\sc ncbu}). The second is where not all fragments are captured, which is termed \emph{incomplete fusion} ({\sc icf}). The final possibility is where the light nucleus is captured completely by the target nucleus and is termed \emph{complete fusion} ({\sc cf}) \cite{Boselli1}. An important aspect is the interplay between breakup and other reaction processes like the transfer process. However due to the transfer process's similarities with the {\sc icf} process it is hard to separate them from an experimental point of view as the fusion products from both processes are the same.  Transfer can also cause the breakup of weakly bound nuclei during low-energy collisions \cite{Vogt1968,Alexis0,Shrivastava2006,Ramin2010,Huy2011,Huy2013,Kalkal1,Cook1}. 

Different types of models have been used to investigate low-energy fusion dynamics of weakly bound nuclei, ranging from classical to quantum mechanical methods. Ref. \cite{Boselli1} provides a critical survey of different theoretical approaches. New studies on the inclusive non-elastic breakup cross section may provide a quantum mechanical route to the calculation of the {\sc icf} cross section of weakly bound nuclei \cite{Hussein1}. Another interesting quantum mechanical framework is the time-dependent wave-packet ({\sc tdwp}) method \cite{Yabana1,Boselli2}. This method calculates the incomplete and complete fusion cross-sections unambiguously \cite{Boselli2}, which is a challenge using the continuum discretized coupled-channels ({\sc cdcc}) method \cite{AlexisThompson0,AlexisThompson1,Thompson2003}. The {\sc tdwp } approach is currently undergoing further development to be implemented using a three-dimensional reaction model. 

Some of the challenges of the quantum mechanical models can be overcome via the use of the three-dimensional classical dynamical model \cite{platypus1,platypus2,platypus3}. This model is implemented using the {\sc platypus} code \cite{platypus3}, which uses classical trajectories in conjunction with stochastic breakup \cite{platypus1,platypus2}. This is done through the input, which includes a breakup function \cite{Hinde0,platypus2}, determined from sub-barrier breakup measurement \cite{Ramin2010,Huy2013}, that undergoes Monte-Carlo sampling \cite{platypus1,platypus2}. This breakup function encodes the effect of the Coulomb and nuclear interactions that cause the breakup, making this approach a quantitative dynamical model for relating the sub-barrier {\sc ncbu} to the above-barrier {\sc icf} and {\sc cf} of weakly bound nuclei, rather than a breakup model \cite{platypus1,platypus2,platypus3}. It is important to note however that this fusion model only works at energies above the Coulomb barrier generated between the projectile and target. This is due to the absence of quantum tunneling that is the primary way of fusion at sub- and near-barrier energies. There has been a recent attempt to amend this classical model by adding a correction at sub- and near-barrier energies, to take into account quantum tunneling. This was done by incorporating a tunneling factor based on the {\sc wkb} approximation \cite{Kharab}. This improved the results outputted from the model, relative to experimental sub-barrier fusion measurements \cite{Kharab}. Additional modifications have been suggested for interpreting sub-barrier breakup measurements \cite{Cook1}.

The dynamics surrounding prompt and delayed breakup is important to understand \cite{Kalkal1}, and this is explored in the present work using {\sc platypus}. For instance, \emph{prompt} breakup happens in the instant the excitation of the $^6$Li projectile is chosen to take place. At this point $^6$Li is converted into its cluster fragments (alpha-deuteron) and then the fragments and target propagate according to the defined interactions between them \cite{platypus2}. \emph{Delayed} breakup is induced by reaching the $1^+$, $2^+$ or $3^+$ resonant states in $^6$Li, which then triggers the dissociation of $^6$Li with the delay coming from the half-life of the resonant $^6$Li state. The $3^+$ resonant state has a much longer half-life than the $2^+$ and $1^+$ resonant states, so the $^6$Li breakup takes place at the outgoing branch of its trajectory, far away from the target nucleus, not affecting fusion. So the effect of the $3^+$ resonance on fusion will be neglected in the calculations below. The $1^+$, $2^+$ or $3^+$ resonant states in $^6$Li can be described by resonant alpha-deuteron $d$-states with energy and width of $(E_{res},\Gamma_{res})\equiv(4.18,1.5)$, $(2.84,1.7)$ and $(0.716,0.024)$ MeV, respectively. Ref. \cite{AlexisThompson1} provides details about the alpha-deuteron continuum states of $^6$Li.

A few required additions to the original version of {\sc platypus} \cite{platypus3} are explained next, followed by a discussion about model calculations for {\sc icf} in $^6$Li + $^{209}$Bi collisions at energies above the Coulomb barrier. This heavy-ion system is a good test case as unambiguous experimental {\sc icf} data exist \cite{Dasgupta1}. These experimental {\sc icf} data refer to the capture of a part of the $^6$Li \emph{charge} by the $^{209}$Bi target \cite{Dasgupta1}, and this definition is adopted in the present study. Conclusions will be drawn about the impact of both the $^6$Li resonances as well as transfer-triggered breakup (involving the projectile-like nuclei $^8$Be and $^5$Li) on the {\sc icf} cross-sections.     
      
\emph{Model and numerical details.} The key additions to the original version of {\sc platypus} \cite{platypus3} are as follows:
\begin{enumerate} 
\item[(i)] A new branching-ratio ($BR$) variable is introduced which helps differentiate between delayed (resonant) and prompt (non-resonant) breakup and it is equal to zero and unity when looking at pure delayed ($BR=0$) and pure prompt ($BR=1$) breakup, respectively. (Only these extreme cases are considered in the present work.) An uniformly generated random number between zero and unity, $n$, is compared with a fixed value of $BR$: if $n > BR$ delayed breakup of an excited projectile happens, provided the sampled excitation energy and intrinsic angular momentum of the projectile \cite{platypus1} correspond to a resonance state characterized by its energy and width, otherwise prompt breakup occurs. 
\item[(ii)] The treatment of lifetime effects in breakup is simpler than the treatment suggested in Refs. \cite{Kalkal1,Cook1}. For instance, for $^5$Li which originates from the n-stripping process, the ground-state (g.s.) is a resonant alpha-proton $p$-state with $(E_{res},\Gamma_{res}) \equiv (1.96,2.0)$ MeV \cite{Jeff1}. If a resonant excitation event is sampled, the original location of the excited projectile on the projectile-target orbit is changed to a new location determined by the resonance half-life $\tau=\hbar/\Gamma_{res}$ that provides additional propagation-time, so the breakup process may occur either closer to the target in the ingoing branch of the projectile-target trajectory or farther away in the outgoing branch of that trajectory. We have also sampled a distribution of time shifts using an exponential decay law ($\propto e^{-t/\tau}$), and the present {\sc icf} cross sections ($BR=0$) increase by $\sim 15\%$.
\item[(iii)] The measured 'breakup triggered by n-stripping and d-pickup' functions contain information about both the production probability of $^5$Li and $^8$Be as well as their breakup probabilities \cite{Huy_private}. These functions take care of how much contribution these channels have as compared to the original $^6$Li channel \cite{Huy_private}, so these functions also describe the competition between the transfer-triggered breakup and the direct breakup of $^6$Li. Consequently, for those transfer channels, $^5$Li and $^8$Be are considered initial projectiles with modified incident energies to match the distance of closest approach $R_{min}$, i.e., $E_0+Q_{tr}$, being $E_0$ and $Q_{tr}$ the $^6$Li incident energies and the average transfer Q-values, respectively. The $Q_{tr}^{gg}$ values are -$1.059$ MeV (n-stripping) and $13.338$ MeV (d-pickup).     
Effects on $R_{min}$ due to uncertainties of $Q_{tr}$ caused by excitations of both projectile-like and target-like nuclei are not included. 
\end{enumerate}        

The measured (prompt) breakup functions deduced from the $Q$-value spectra of the studied channels \cite{Ramin2010,Huy2013} are characterized by two constants, ($\alpha$,$\beta$), of the exponential function $P_{BU}(R_{min})=\exp(\alpha R_{min} + \beta)$. These values are $(-0.353738,1.26228)$, $(-0.799239,8.03581)$ and $(-1.57787,14.3403)$ for $^6$Li + $^{209}$Bi, $^5$Li + $^{210}$Bi and $^8$Be + $^{207}$Pb systems, respectively \cite{Huy_private}. Refs. \cite{Huy2013,Kalkal1,Cook1} explain how these functions are experimentally determined. 

The parameters of the Woods-Saxon nuclear interaction between the projectile fragments and the target are determined by approximately matching the corresponding Sao-Paulo potential barriers \cite{SP}. Table \ref{table:pot} presents those parameters, while for the Coulomb interaction the potential of a uniformly charged sphere has been used. All the radius parameters provide a distance determined by $r_0 A^{1/3}$, where $A$ is the heaviest mass in the corresponding binary system. 

\begin{table}[ht]
\caption{Parameters of the Woods-Saxon nuclear potential for different systems in the present calculations as well as the radius parameter of their Coulomb interactions (last column).}
\centering
\begin{tabular}{c c c c c}
\hline\hline
Systems & $V_0$ (MeV) & $r_0$ (fm) & $a_0$ (fm) & $r_{0c}$ (fm) \\
\hline 
$^{209}$Bi + $^6$Li & -51.761 & 1.545 & 0.659 & 1.2 \\
$^{209}$Bi + $^4$He & -32.931 & 1.461 & 0.605 & 1.2 \\
$^{209}$Bi + $^2$H  & -26.000 & 1.465 & 0.668 & 1.2 \\
$^4$He + $^2$H      & -78.460 & 1.150 & 0.700 & 1.465 \\
$^{210}$Bi + $^5$Li & -51.761 & 1.513 & 0.663 & 1.2 \\ 
$^{210}$Bi + $^4$He & -32.931 & 1.459 & 0.608 & 1.2 \\ 
$^{210}$Bi + $^1$H  &  -9.900 & 1.320 & 0.679 & 1.2 \\
$^4$He + $^1$H      & -52.350 & 1.100 & 0.378 & 1.2 \\
$^{207}$Pb + $^8$Be & -120.903 & 1.430 & 0.762 & 1.2 \\ 
$^{207}$Pb + $^4$He & -62.000 & 1.385 & 0.620 & 1.2 \\
$^4$He + $^4$He & -16.696 & 1.200 & 0.620 & 1.2 \\ [0.5ex]
\hline
\end{tabular}
\label{table:pot}
\end{table}

Table \ref{table:gauss} presents the centroid ($d_{012}$) and variance ($\sigma_{012}$) of the Gaussian distributions for the radial g.s. probability density of the different two-body projectiles \cite{platypus3}. These parameters describe the radial probability density inside the radius of the Coulomb barrier between the fragments. This radial probability density is derived from the projectile g.s. wave-function. In sampling breakup, the maximal internal energy and angular momentum of the two-body projectiles are $\epsilon_{max}=6$ MeV and $l_{max}=4\hbar$, as in Ref. \cite{platypus1}. Resonant breakup events are also sampled in this energy and angular momentum window, as explained in point (i). Prompt breakup for partial waves other than a resonant wave is included in the energy region of a resonance. Only even $l$ are included when the breakup fragments are identical. These $\epsilon_{max}$ and $l_{max}$ values as well as the values of the orbital angular momenta of the projectiles ($L_0 \leq 100\hbar$) guarantee the convergence of the above-barrier {\sc icf} and {\sc ncbu} cross-sections.

\begin{table}[ht]
\caption{Centroid and variance of the Gaussian distributions for the radial g.s. probability density of the different two-body projectiles.}
\centering
\begin{tabular}{c c c}
\hline\hline
Projectiles & $d_{012}$ (fm) & \, $\sigma_{012}$ (fm) \\ [0.5ex]
\hline
$^6$Li\,($^4$He + $^2$H) & 3.12 & 1.05 \\
$^5$Li\,($^4$He + $^1$H) & 2.95 & 1.78 \\
$^8$Be\,($^4$He + $^4$He) & 1.80 & 1.00 \\ [1ex]
\hline
\end{tabular}
\label{table:gauss}
\end{table}

\begin{figure}
\centering
\includegraphics[width=8.5cm,clip]{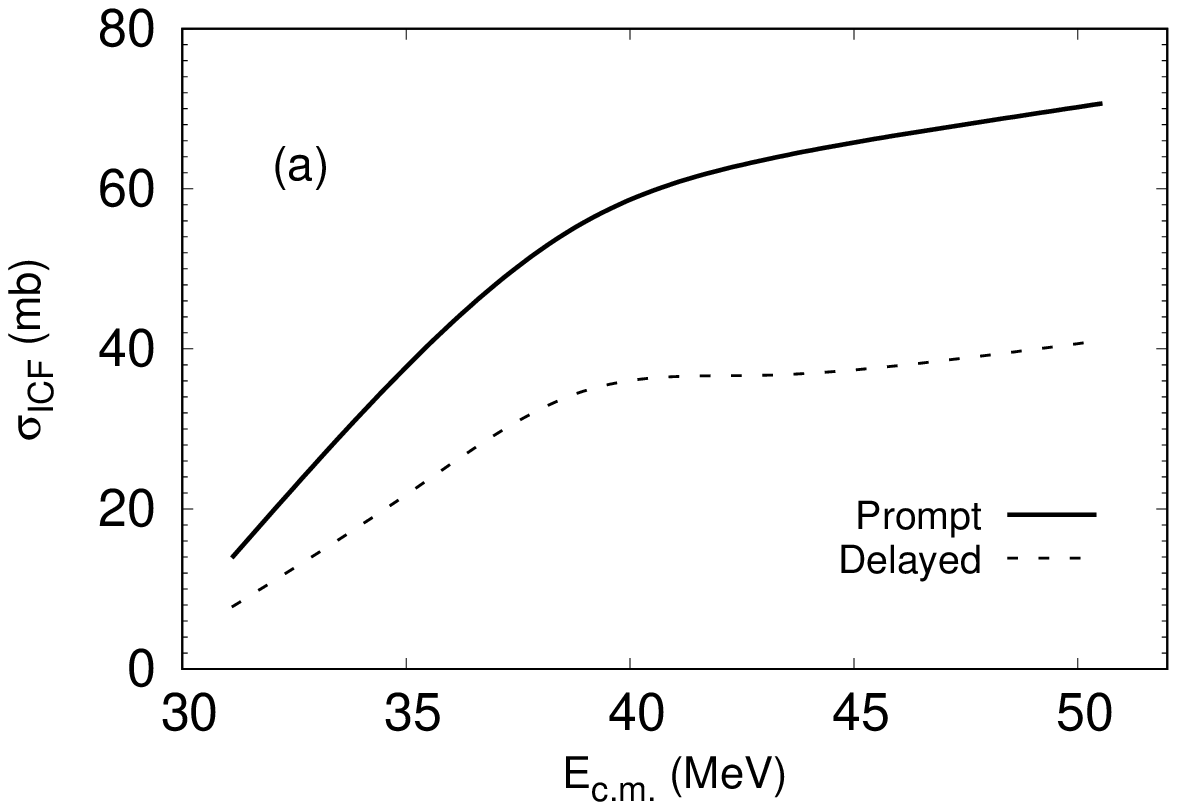} \\
\includegraphics[width=8.5cm,clip]{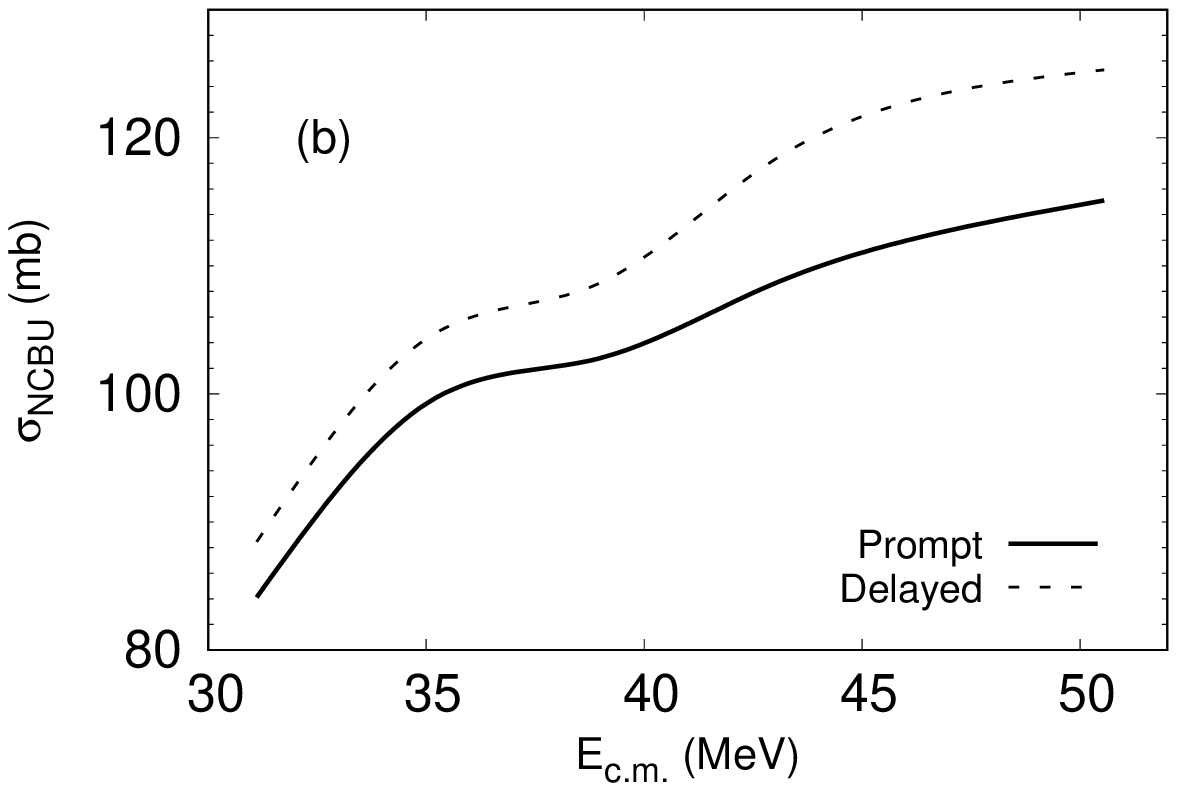}
\caption{(a) Incomplete-fusion excitation function when direct breakup of $^{6}$Li into alpha and deuteron happens in collisions with the $^{209}$Bi target: prompt breakup (solid line), delayed breakup (dashed line). (b) The same but for the no-capture breakup excitation curve.}
\label{fig-2}       
\end{figure}

\emph{Results.}  Figure \ref{fig-2}(a) shows a clear difference between the {\sc icf} cross-sections for delayed (dashed line) and prompt (solid line) direct breakup of $^6$Li. The lower cross-sections for delayed breakup can be explained by the fact that when the projectile is excited to a resonant state the projectile flies past the target, due to the half-life associated with the resonant state, before it breaks up. Most breakup events happen close to the distance of minimal approach \cite{platypus1,platypus2}. The half-lives for the $1^+$ and $2^+$ states are of $4.2 \times 10^{-22}$s and $3.7 \times 10^{-22}$s, respectively. The typical collision time is $\sim 10^{-21}$s, therefore these short-lived states can have a significant effect on the {\sc icf} process. This causes the probability of an {\sc icf} reaction taking place to decrease, as it is less likely for a fragment to be absorbed when resonance states ($1^+$, $2^+$) are excited near the target. Consequently, the {\sc icf} and {\sc ncbu} cross-sections are anticorrelated as observed in Fig. \ref{fig-2}(b). When the projectile is excited to a resonant state, the projectile breaks up in the outgoing branch of its trajectory, making it less likely for a fragment to be absorbed by the target.

\begin{figure}
\centering
\includegraphics[width=8.5cm,clip]{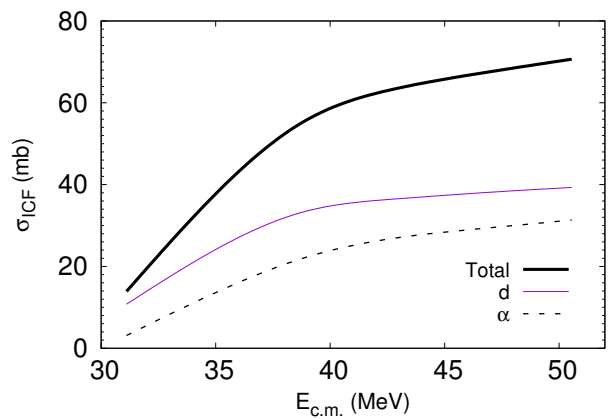}
\caption{(Color online) Alpha and deuteron contributions to the incomplete-fusion excitation function for prompt direct breakup of $^{6}$Li in collisions with $^{209}$Bi.}
\label{fig-3}       
\end{figure}

\begin{figure}
\centering
\includegraphics[width=8.5cm,clip]{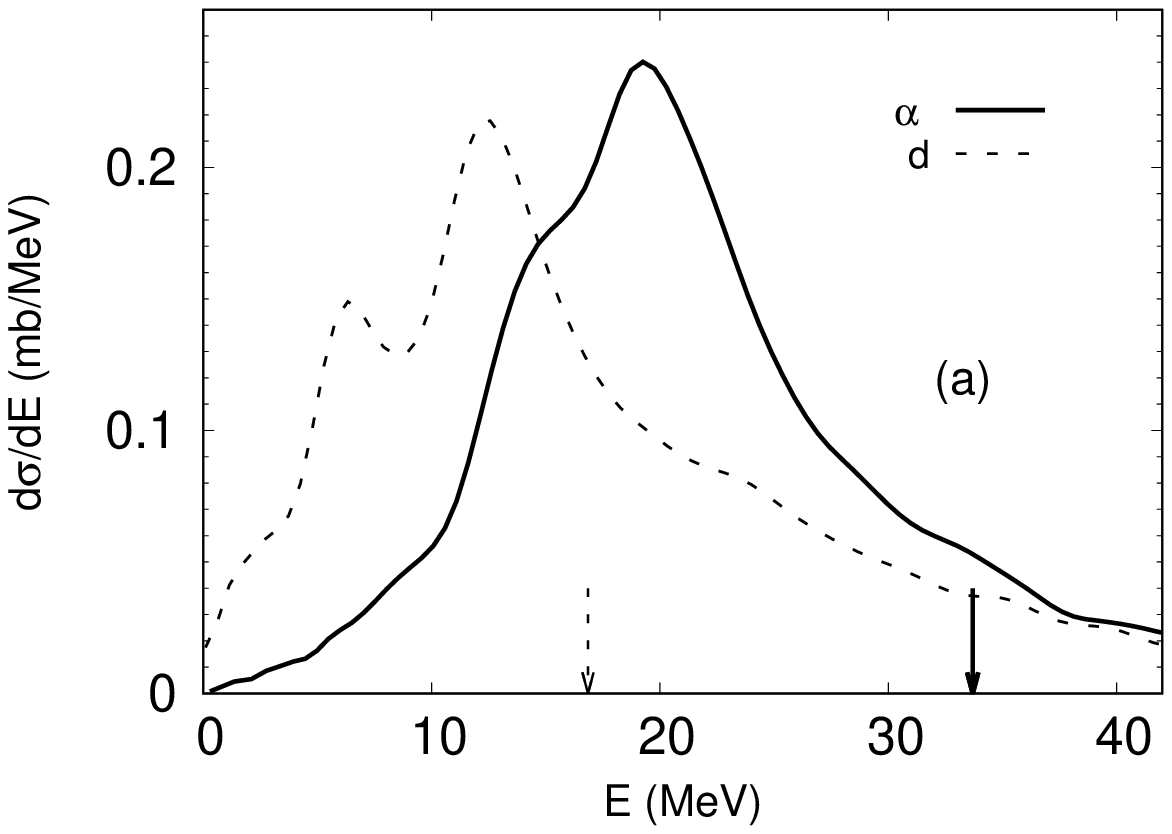} \\
\includegraphics[width=8.5cm,clip]{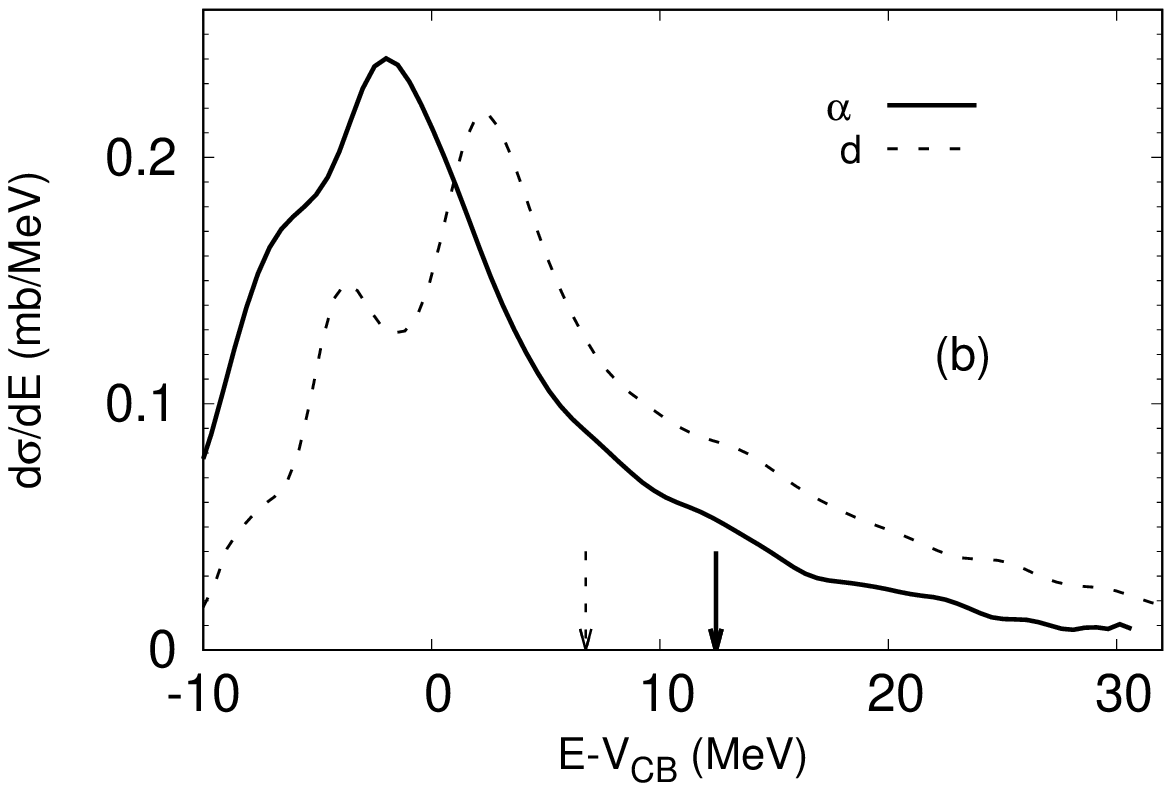}
\caption{(a) Relative kinetic-energy distributions for $\alpha-^{209}$Bi and $d-^{209}$Bi immediately following the prompt direct breakup of $^{6}$Li in collisions with $^{209}$Bi at $E_{c.m.}=50.5$ MeV. (b) The same but the energy is measured relative to the height of the Coulomb barriers between each cluster and the $^{209}$Bi target. The arrows denote the partition of the $^{6}$Li incident energy between these clusters according to their masses.}
\label{fig-3new}       
\end{figure}

\begin{figure}
\centering
\includegraphics[width=8.5cm,clip]{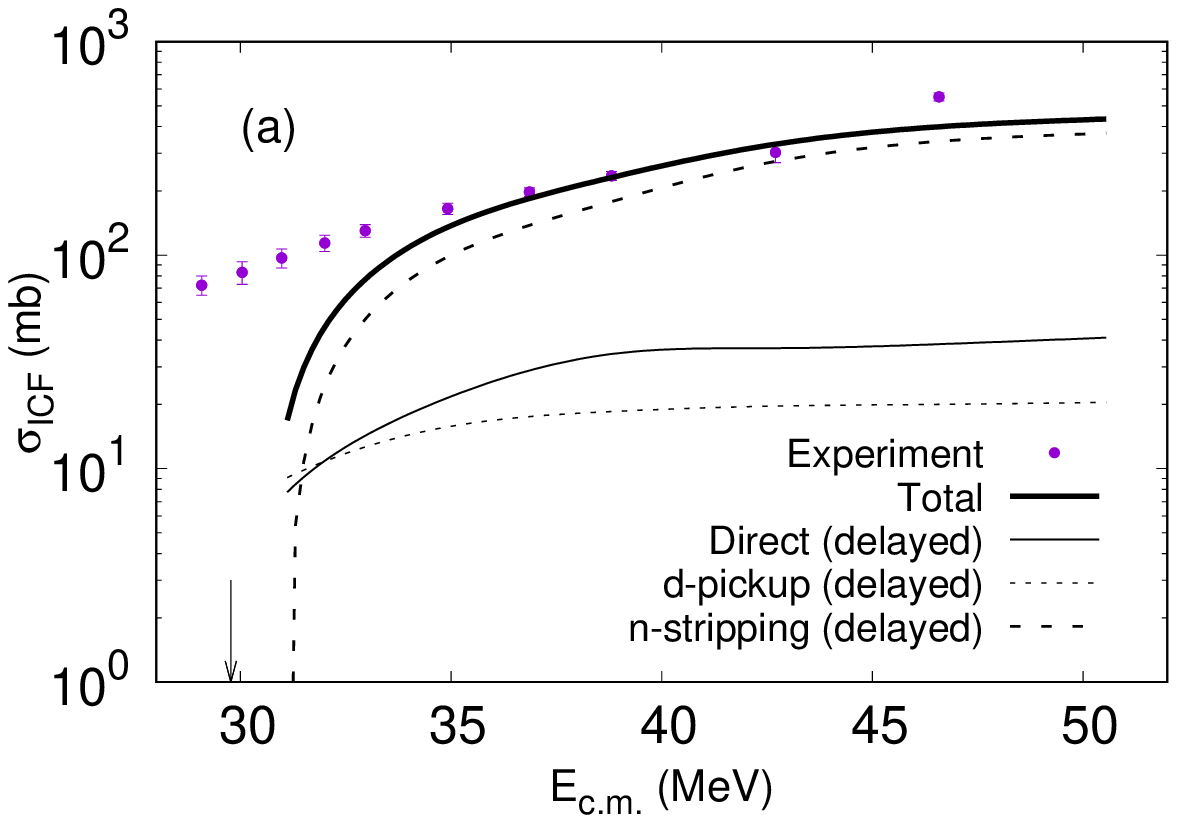} \\
\includegraphics[width=8.5cm,clip]{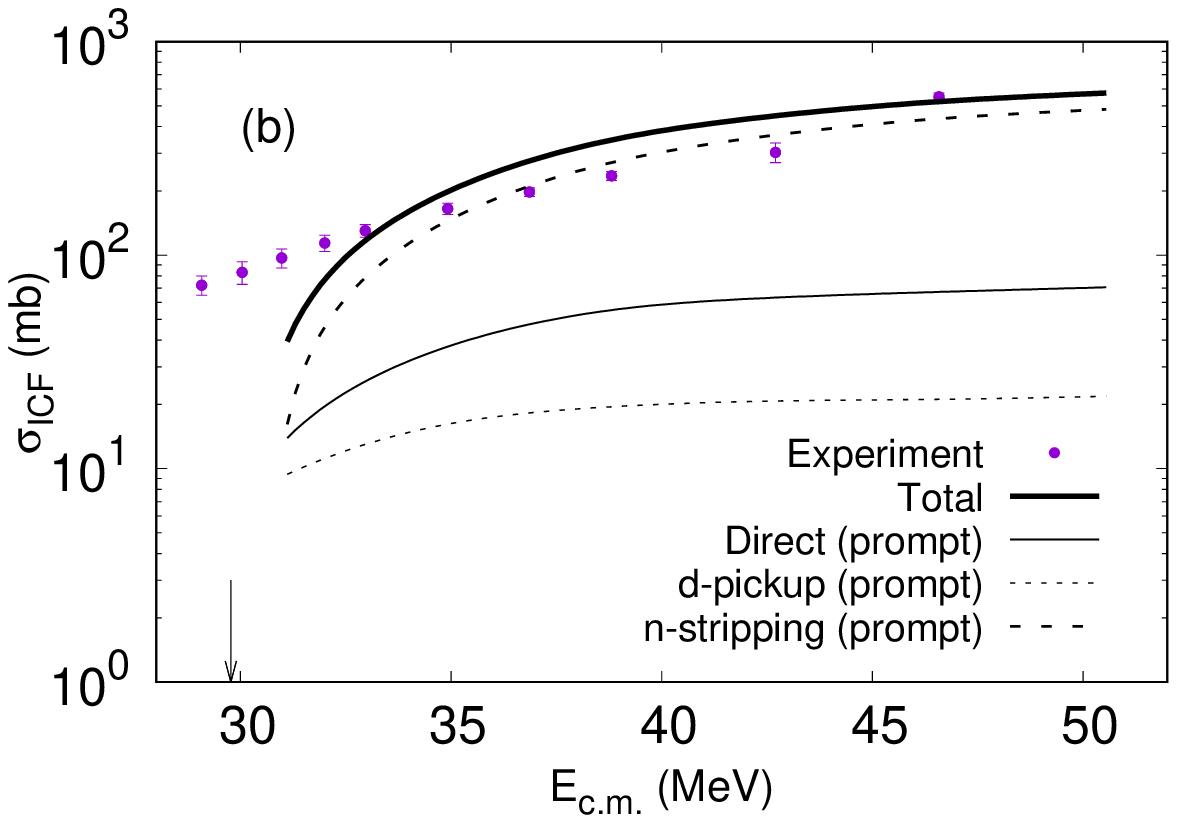}
\caption{(Color online) Experimental incomplete-fusion cross-sections for $^6$Li + $^{209}$Bi \cite{Dasgupta1} are compared with {\sc platypus} calculations at above-barrier (arrow) energies. Direct breakup as well as transfer-triggered breakup channels are included. (a) For \emph{delayed} direct breakup of $^6$Li, \emph{delayed} breakup of $^8$Be after d-pickup, and \emph{delayed} breakup of $^5$Li after n-stripping. (b) The same but for \emph{prompt} breakup processes.}
\label{fig-4}       
\end{figure}

Figure \ref{fig-3} shows the alpha (dashed line) and deuteron (thin solid line) contributions to the direct {\sc icf} excitation function (thick solid line). The deuteron contribution is higher than the alpha contribution due to the respective difference in Coulomb barriers between the alpha and the target ($\sim 21.25$ MeV) and the deuteron and the target ($\sim 10.1$ MeV). For instance, Fig. \ref{fig-3new} shows, for the {\sc icf} process, the relative kinetic energy distributions for $\alpha-^{209}$Bi and $d-^{209}$Bi immediately following the prompt direct breakup of $^6$Li at the incident energy of $E_{c.m.}=50.5$ MeV. Although the position of the maxima of these distributions qualitatively agrees with a simple partition of the $^6$Li incident energy between the clusters according to their masses (arrows), as depicted in Fig. \ref{fig-3new}(a), the role of the individual Coulomb barriers in the respective {\sc icf} process is crucial, as observed in Fig. \ref{fig-3new}(b). In Fig. \ref{fig-3new}(b), we can see that the yield of deuterons with a positive kinetic energy (dashed line) is substantially larger than that for alpha particles (solid line), explaining the alpha and deuteron contributions in Fig. \ref{fig-3}. The trends in Fig. \ref{fig-3} are also observed in the simplified {\sc tdwp} calculations discussed in Ref. \cite{Boselli2}, and disagree with those in Ref. \cite{kailas2016} where the {\sc icf} treatment neglects the competition and correlation between the alpha- and deuteron-capture processes. In Ref. \cite{kailas2016} it is assumed that all the observed {\sc icf} products \cite{Dasgupta1} (i.e., actinium and polonium isotopes) were originated from the $^6$Li direct breakup process, which is neither the main observed breakup channel \cite{Huy2013,Kalkal1,Cook1} nor the dominant {\sc icf} route as demonstrated below.  

Figure \ref{fig-4} presents the different contributions of various reaction processes to the total {\sc icf} cross-sections (thick solid line) which are compared with experimental data \cite{Dasgupta1}. Together with the direct (delayed and prompt) breakup of $^6$Li, transfer-triggered breakup of projectile-like nuclei such as $^8$Be (after d-pickup) and $^5$Li (after n-stripping) occurs during collisions of $^6$Li and $^{209}$Bi at energies near the Coulomb barrier. These projectile-like nuclei are unstable and their short-lived states may affect the formation of {\sc icf} products. Only prompt breakup of $^8$Be excited states can impact on the {\sc icf} cross-sections as its g.s. half-life is very long ($\sim 10^{-16}$s) compared to the collision time. The $^8$Be g.s. breakup has been neglected, but both its prompt breakup and delayed breakup via its first $2^+$ resonant state with $(E_{res},\Gamma_{res}) \equiv (3.0,1.5)$ MeV \cite{8Be} are included. In contrast to $^8$Be, $^5$Li has a short-lived g.s. ($\sim 3.3 \times 10^{-22}$s) whose decay can substantially affect the {\sc icf} cross-sections as shown in Fig. \ref{fig-4}(a) (dashed line). The n-stripping channel provides the dominant contribution to the {\sc icf} cross-sections. Figs. \ref{fig-4}(a) and \ref{fig-4}(b) depict extreme cases of breakup (delayed and prompt), so the real scenario is somewhere in between, which reasonably agrees with the observations. The present classical dynamical model does not treat quantum tunneling, so the description of the experimental data at energies very close to the Coulomb barrier ($\sim 29.8$ MeV) is not reliable.        

\emph{Conclusions.} The incomplete fusion process for $^6$Li + $^{209}$Bi collisions at energies above the Coulomb barrier has been investigated with the classical dynamical model implemented in the {\sc platypus} code. The main conclusions of the present work are as follows:
\begin{enumerate}
\item The resonant states ($1^+$, $2^+$) of $^6$Li play an important role in the direct {\sc icf} and {\sc ncbu} cross-sections, due to the respective half-life of these resonant states ($\sim 10^{-22}$s) relative to the collision time ($\sim 10^{-21}$s). Delayed breakup via excitation of those resonance states reduce the theoretical {\sc icf} cross-sections.
\item The deuteron contribution to the direct component of {\sc icf} cross-sections (this component is much smaller than the n-stripping component) is significantly higher than the alpha contribution because of the much smaller Coulomb barrier between the deuteron and the $^{209}$Bi target (by $\sim 11$ MeV).
\item The n-stripping channel involving the projectile-like nucleus $^5$Li clearly dominates the formation of {\sc icf} products. This is the central result of the present work. In contrast most quantum mechanical fusion calculations assume that the direct breakup of $^{6}$Li is the dominant  {\sc icf} channel. 
\item Using information from sub-barrier breakup measurements, {\sc platypus} provides a comprehensive and insightful explanation of the {\sc icf} excitation function at above-barrier energies.
\end{enumerate}

\acknowledgments
The support from the STFC grant (ST/P00671X/1) is acknowledged. AD-T thanks both J.A. Tostevin for calculations on the $^5$Li g.s. properties and D.H. Luong for providing information on the experimental breakup functions.


\begin{thebibliography}{99}

\bibitem{Wiescher1} M. Wiescher, F. K\"appeler and K. Langanke, Ann. Rev. of Astron. and Astrophys. \textbf{50}, 165 (2012).

\bibitem{Back1} B.B. Back, H. Esbensen, C.L. Jiang and K.E. Rehm, Rev. Mod. Phys. \textbf{86}, 317 (2014).

\bibitem{Canto1} L.F. Canto, P.R.S. Gomes, R. Donangelo and M.S. Hussein, Phys. Rep. \textbf{596}, 1 (2015). 

\bibitem{Boselli1} M. Boselli and A. Diaz-Torres, J. of Phys. G \textbf{41}, 094001 (2014).

\bibitem{Vogt1968} M.A. Reimann, P.W. Martin and E.W. Vogt, Can. J. of Phys. \textbf{46}, 2241 (1968).

\bibitem{Alexis0} A. Diaz-Torres, I.J. Thompson and W. Scheid, Phys. Lett. B \textbf{533}, 265 (2002); Nucl. Phys. A \textbf{703}, 83 (2002).

\bibitem{Shrivastava2006} A. Shrivastava et al., Phys. Lett. B \textbf{633}, 463 (2006).

\bibitem{Ramin2010} R. Ramin et al., Phys. Rev. C \textbf{81}, 024601 (2010).

\bibitem{Huy2011} D.H. Luong et al., Phys. Lett. B \textbf{695}, 105 (2011).

\bibitem{Huy2013} D.H. Luong et al., Phys. Rev. C \textbf{88}, 034609 (2013).

\bibitem{Kalkal1} S. Kalkal et al., Phys. Rev. C \textbf{93}, 044605 (2016).

\bibitem{Cook1} K.J. Cook et al., Phys. Rev. C \textbf{93}, 064604 (2016).

\bibitem{Hussein1} B.V. Carlson, T. Frederico and M.S. Hussein, Phys. Lett. B \textbf{767}, 53 (2017).

\bibitem{Yabana1} K. Yabana, Prog. Theor. Phys. \textbf{97}, 437 (1997).

\bibitem{Boselli2} M. Boselli and A. Diaz-Torres, Phys. Rev. C \textbf{92}, 044610 (2015).

\bibitem{AlexisThompson0} A. Diaz-Torres and I.J. Thompson, Phys. Rev. C \textbf{65}, 024606 (2002).

\bibitem{AlexisThompson1} A. Diaz-Torres, I.J. Thompson and C. Beck, Phys. Rev. C \textbf{68}, 044607 (2003).

\bibitem{Thompson2003} I.J. Thompson and A. Diaz-Torres, Prog. Theor. Phys. Suppl. \textbf{154}, 69 (2004). 

\bibitem{platypus1} A. Diaz-Torres, D.J. Hinde, J.A. Tostevin, M. Dasgupta and L.R. Gasques, Phys. Rev. Lett. \textbf{98}, 152701 (2007).

\bibitem{Hinde0} D.J. Hinde et al., Phys. Rev. Lett. \textbf{89}, 272701 (2002).

\bibitem{platypus2} A. Diaz-Torres, J. of Phys. G \textbf{37}, 075109 (2010).

\bibitem{platypus3} A. Diaz-Torres, Comp. Phys. Comm. \textbf{182}, 1100 (2011).

\bibitem{Kharab} R. Kharab, R. Chahal and R. Kumar, Nucl. Phys. A \textbf{960}, 11 (2017).


\bibitem{Dasgupta1} M. Dasgupta et al., Phys. Rev. C \textbf{70}, 024606 (2004).

\bibitem{Jeff1} V.D. Efros and H. Oberhummer, Phys. Rev. C \textbf{54}, 1485 (1996); J.A. Tostevin (private communication).

\bibitem{Huy_private} D.H. Luong (private communication).

\bibitem{SP} L.C. Chamon et al., Phys. Rev. C \textbf{66}, 014610 (2002).

\bibitem{kailas2016} V.V. Parkar, V. Jha and S. Kailas, Phys. Rev. C \textbf{94}, 024609 (2016).

\bibitem{8Be} https://www.nndc.bnl.gov 


\end{thebibliography}
\end{document}